\documentclass{iopart}

\usepackage{graphicx}
\begin{document}

\title[Charmonium dissociation temperatures in lattice QCD]
{Charmonium dissociation temperatures in lattice QCD with a finite
volume technique} 

\author{T.~Umeda\footnote{speaker}, H.~Ohno, 
K.~Kanaya for the WHOT-QCD Collaboration}
\address{Graduate School of Pure and Applied Sciences,
University of Tsukuba, Tsukuba, Ibaraki 305-8571, Japan}
\ead{tumeda@het.ph.tsukuba.ac.jp}

\begin{abstract}
Dissociation temperatures of $J/\psi$,  $\psi'$, and $\chi_c$
states play key roles in the sequential $J/\psi$ suppression scenario 
for high energy heavy ion collisions.
We report on a study of charmonium dissociation temperatures in
quenched lattice QCD. 
On anisotropic lattices, we first subtract the effects of the constant
mode in finite temperature meson correlators, which have lead to
unphysical results in previous studies. 
We then extract ground and first exited state masses by diagonalizing
correlation functions among different source and sink operators.
To distinguish bound states from scattering states, we first compare
the charmonium mass spectra under different spatial boundary
conditions, and examine the shape and the volume-dependence of their Bethe-Salpeter wave
functions. 
From these studies, we found so far no sign of scattering states up to
about $2.3T_c$. 
\end{abstract}


\section{Introduction}

Heavy quarkona play important roles in the study of quark gluon plasma
(QGP) in high energy heavy ion collisions.  
Lattice studies of charmonium spectral functions at finite
temperatures have  
suggested that hadronic excitations corresponding to $J/\psi$ may
survive 
in the deconfinement phase till relatively high temperatures
\cite{Umeda:2000ym,Umeda:2002vr,Datta:2003ww,Asakawa:2003re,Aarts:2006nr,
Jakovac:2006sf}. 
In order to understand discrepancy with the experimental observation
of $J/\psi$ suppression, 
a sequential $J/\psi$ suppression scenario has been proposed 
\cite{Digal:2001bh,Karsch:2005nk}, in which 
a part of $J/\psi$ particles in heavy ion collisions are produced
through $\psi'$ and $\chi_c$'s  
and thus 
dissociation of $\psi'$ and $\chi_c$'s states lead to additional
contributions to   
the $J/\psi$ suppression.

To examine the scenario, 
several groups started to study spectral functions for
$\chi_c$ states on the lattice adopting a Bayesian method
\cite{Datta:2003ww,Aarts:2006nr,Jakovac:2006sf}. 
They found that the low energy part of the spectral functions for
$\chi_{c0}$ and  
$\chi_{c1}$ states drastically increase just above $T_c$, 
suggesting dissociation of these states immediately above $T_c$.
On the other hand, one of the present authors has recently pointed out
that such change is caused by a 
constant mode expected in high temperature meson correlation functions
on the lattice \cite{Umeda:2007hy}. 
To obtain physically meaningful results from meson correlation
functions in the deconfining phase, we have to separate the effects of
the constant mode. 

In this paper, we carry out a study of charmonium states above $T_c$
in quenched QCD on anisotropic lattices, subtracting the contribution of
the constant mode.

\section{Our approach}
\label{sec:approach}
We combine all the known rigid techniques to investigate meson states at
high temperature. 
In order to extract a few lowest energy states in Euclidean correlators
we adopt the variational method \cite{Luscher:1990ck}, 
in which the eigenvalues
$\lambda_\alpha(t,t_0)$ of the correlation matrix 
provide the energies of the states. 
\begin{eqnarray}
X^{-1}{C(t_0)}^{-1}C(t)X&=&diag\left\{\lambda_\alpha(t,t_0)
\right\},~~~(\alpha=1,\cdots,N).\label{eq001}
\end{eqnarray}
The $N \times N$ correlation matrix is
defined by
\begin{eqnarray}
C_{ij}(t)&=&\sum_{\vec{x}}\langle O_i(\vec{x},t)O^\dagger_j(\vec{0},0)
\rangle,\\
O_i(\vec{x},t)&=&\sum_{y,z}\phi_i(\vec{y})\phi_i(\vec{z})
\bar{q}(\vec{x}+\vec{y},t)\Gamma q(\vec{x}+\vec{z},t)
\end{eqnarray}
where $\Gamma = \gamma_1$ and $\gamma_1\gamma_5$ for 
vector(V) and axial-vector(Av) channels, respectively.
In this study we use operators with Gaussian smearing 
function $\phi_i(\vec{x})=\exp(-A|\vec{x}|^2)$ 
with $N$ different values of the width parameter $A$. 
The analysis enables us to study excited states such as 
$\psi(2S)$, which is the radial
excitation state of charmonium with $J^{PC}=1^{--}$ channel.

As mentioned in the Introduction, 
to study meson-like states in the high temperature phase, we have to subtract the effects of the constant mode from the meson correlators, because it sometimes makes it difficult to extract low energy states, in particular for P-waves.
We adopt the midpoint subtraction method \cite{Umeda:2007hy} to remove the constant mode.

To distinguish bound states of $c$-$\bar{c}$ quarks
from their scattering states in lattice QCD, we study the 
spatial boundary condition dependence of the
energy spectrum at finite volume\cite{Iida:2006mv}:
The energy of scattering states depends on its relative momentum
which is quantized according to the volume size and boundary conditions.
On the other hand the spectrum of the bound states does not change
against such exchange of boundary conditions.
Therefore, we can judge the existence of scattering states from a shift in the energy spectrum.
In this study, we prepare three types of spatial boundary condition, PBC, APBC and MBC, where
(A)PBC is the (anti-)periodic boundary condition in all spatial
directions, and MBC is anti-periodic in x-direction and periodic in
the others directions. 

We also study the Bethe-Salpeter wave function from 
the spatial correlation function between $c$-$\bar{c}$ quarks \cite{Umeda:2000ym}.
We expect that wave functions for binding state will be compact, while those for scattering state will be extended and will change its shape depending on the lattice volume.

\section{Numerical Results}

\subsection{Lattice setup}
\label{sec:setup}
The gauge configurations have been generated by an standard plaquette
gauge action with a lattice gauge coupling constant, $\beta=6.10$ and a
bare anisotropy parameter $\gamma_G=3.2108$. 
The definition of the action and parameters
are the same as that adopted in Ref.~\cite{Matsufuru:2001cp}.
The lattice spacings are $1/a_s=2.030(13)$ GeV and
$1/a_t=8.12(5)$ GeV.
Our simulations are performed on $N_s^3\times N_t$ lattices
with $N_t=32-12$ 
and $N_s=16-32$ to study temperature and volume dependence
respectively.
Temperatures are given by $T/T_c=N_t^{crit}/N_t$ where 
$N_t^{crit}\sim 28$ in our setup.
For the quark fields, we adopt an $O(a)$-improved Wilson quark action
with tadpole improved tree level clover coefficients.
Although the definition of the quark action is the
same as in Ref.~\cite{Matsufuru:2001cp}, 
we adopt a different choice of the Wilson
parameter $r=1$ to suppress lattice artifact in higher excited states.

\subsection{Boundary condition dependences}
\label{sec:bc}

\begin{figure}
\begin{center}
 \includegraphics[width=50mm]{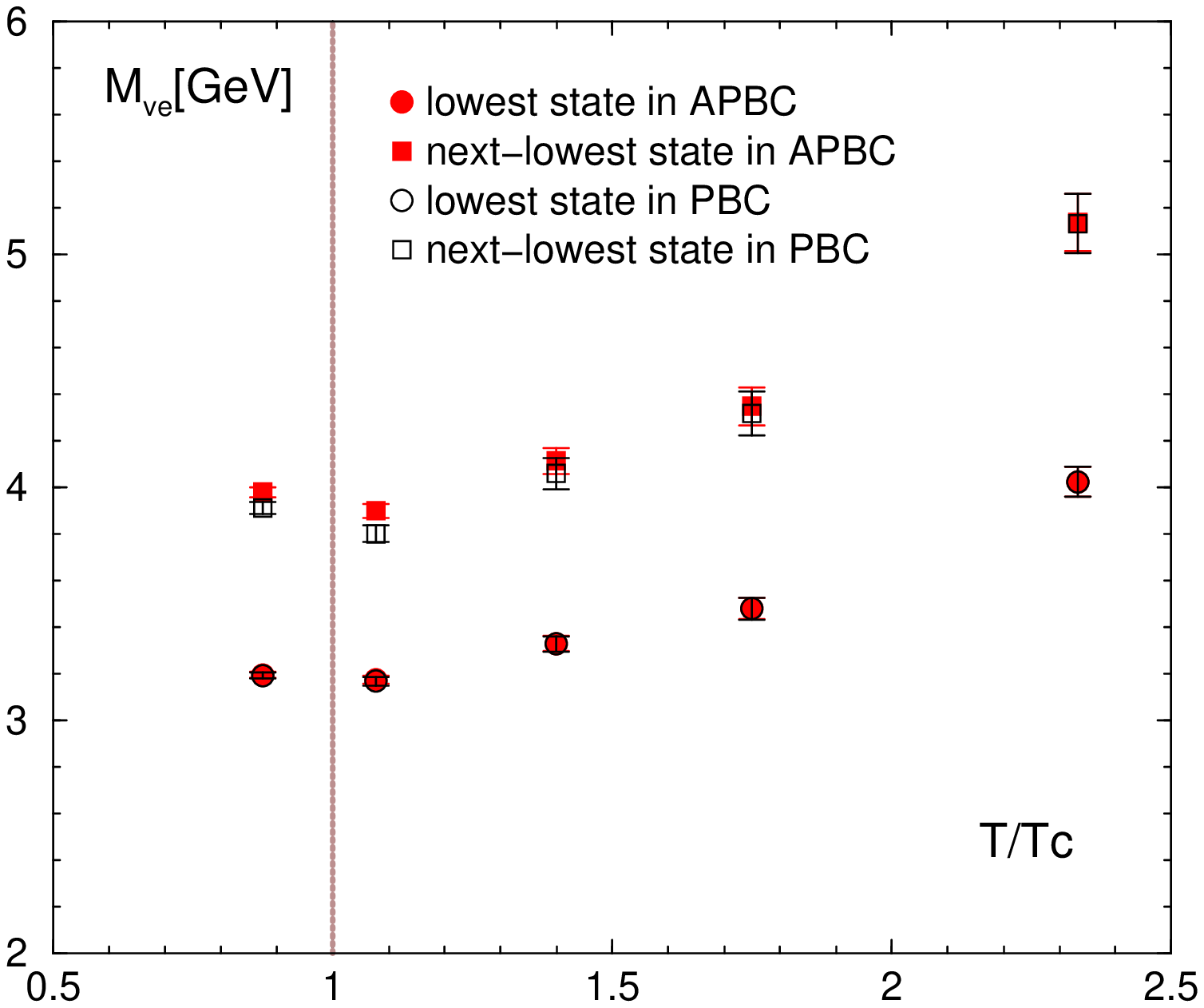}
 \hspace{10mm}
 \includegraphics[width=50mm]{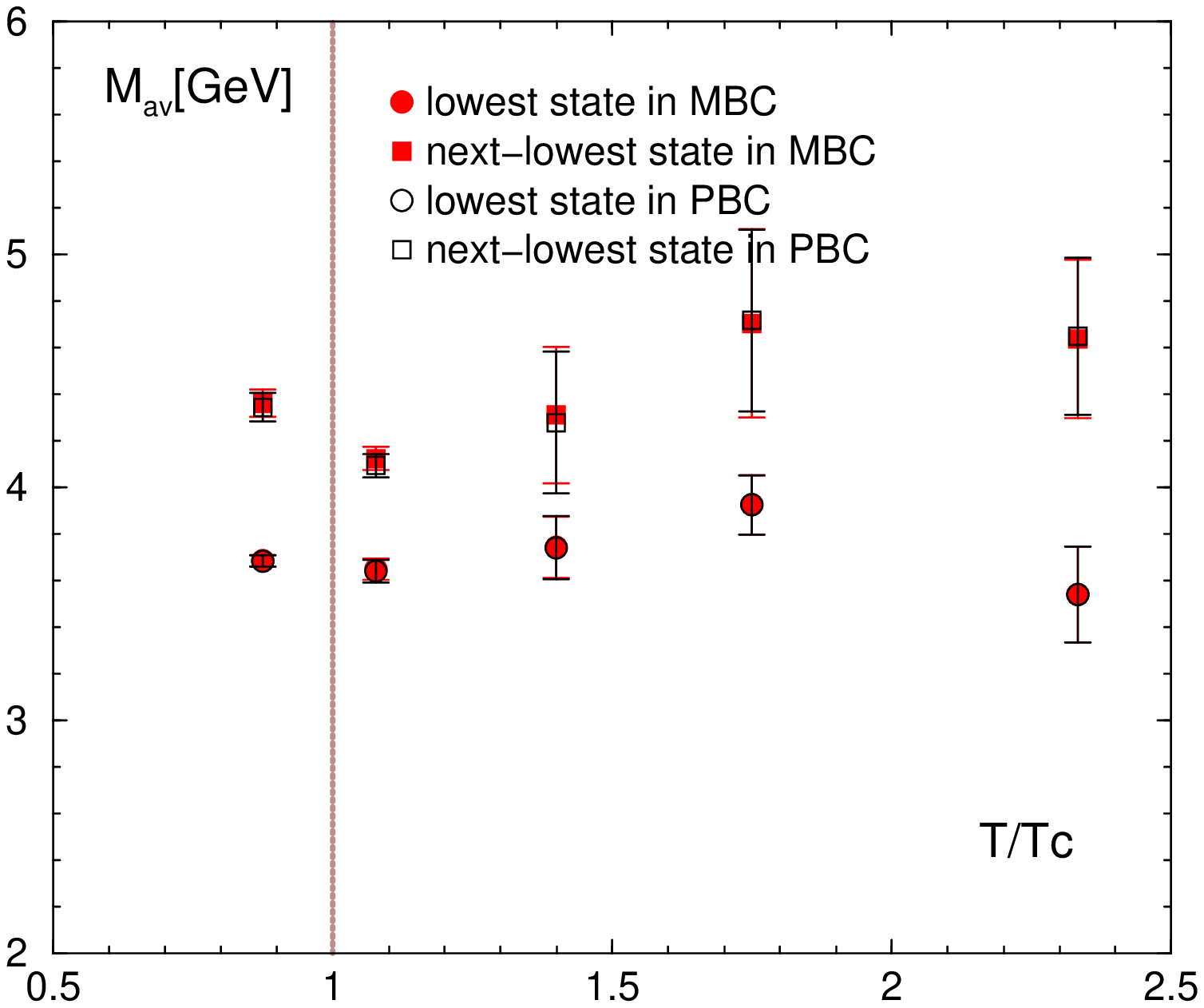}
\caption{Temperature dependences of the lowest and next-lowest state
energies in different boundary conditions. Left and right panels show
charmonium with vector and axial-vector channels respectively.\vspace{-5mm}}
\label{fig:mass}
\end{center}
\end{figure}

Figure \ref{fig:mass} shows the temperature dependence of charmonium
energies in vector and axial-vector channels.
These energies are calculated by the variational method with 
$4 \times 4$ correlation matrix.
At $T=0$ the lowest and next-lowest states in the vector channel are
$J/\psi$ and $\psi(2S)$ states respectively.
The lowest state in the axial-vector channel is the $\chi_{c1}$ state
at $T=0$. 
Results adopting different boundary conditions are shown by different
colors. 

From a study of meson-like correlators in the free quark case as a
model of scattering states, we find that the largest shift of the ground
state energy appears between PBC and APBC for the S-wave, while for P-wave
states between PBC and MBC. 
Therefore, we compare  PBC and APBC(MBC) for S-(P-)waves. 
As a reference, energy shifts in the lowest states are about 200MeV for the case
of free quarks in a $(2fm)^3$ box.

Although statistical errors in axial-vector channels are not quite small yet,
Fig.~\ref{fig:mass} shows that typical mass shifts are much smaller
than those expected in the free quark case up to about $2.3T_c$.

\subsection{Wave functions}
\label{sec:wf}

\begin{figure}
\begin{center}
 \includegraphics[width=55mm]{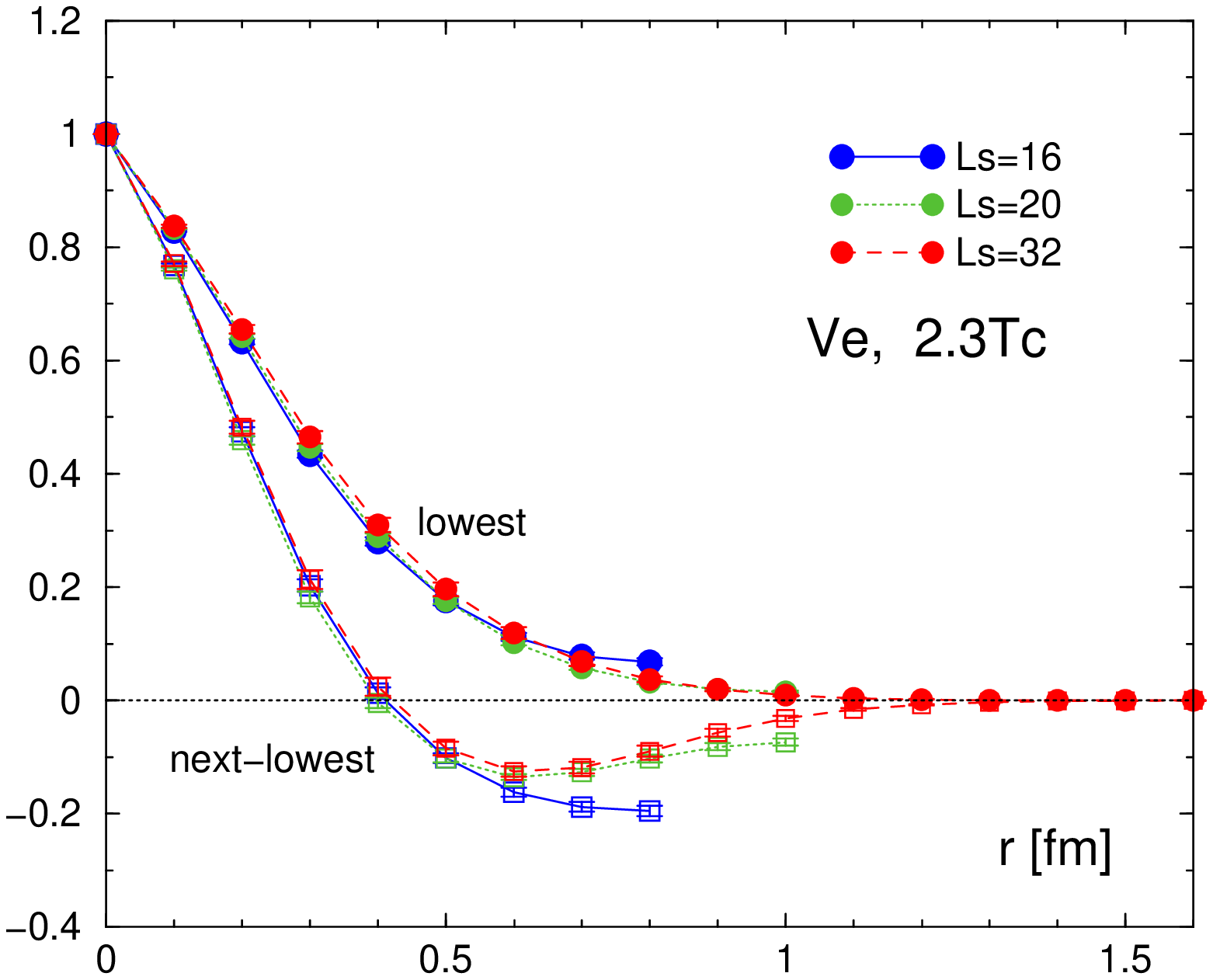}
\hspace{5mm}
 \includegraphics[width=55mm]{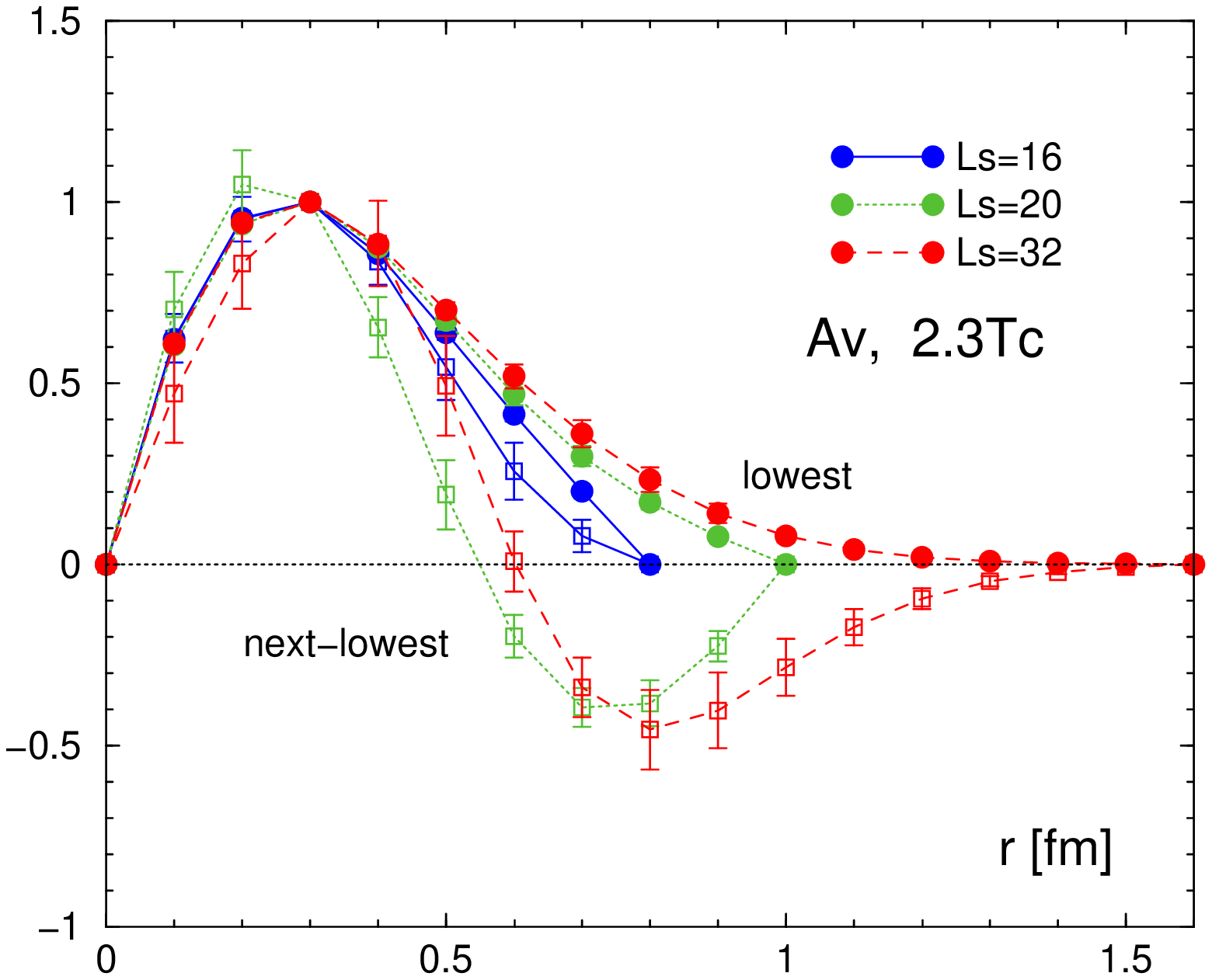}
\caption{Volume dependences of the lowest and next-lowest state
BS wave functions at $T=2.3T_c$. Left and right panels show
charmonium with vector and axial-vector channels respectively.\vspace{-5mm}}
\label{fig:wf}
\end{center}
\end{figure}

Figure \ref{fig:wf} shows the volume dependence of charmonium wave
functions in vector and axial-vector channels at $T=2.3T_c$. 
The Bethe-Salpeter wave function is defined by
\begin{eqnarray}
BS_\alpha(\vec{r},t)&=&\sum_{\vec{x}}\langle
\bar{q}(\vec{x}+\vec{r},t)\Gamma q(\vec{x},t)
|\Omega_\alpha\rangle, \\
\Phi(\vec{r},t) &=& BS_\alpha(\vec{r},t)/BS_\alpha(\vec{r_0},t).
\end{eqnarray}
Here the axial-vector channel is calculated with a derivative operator
$\Gamma=\sum_{jk}( \epsilon_{ijk} \gamma_j \partial_k )$ 
in order to construct the wave
function with large component.
The $|\Omega_\alpha\rangle$ is $\alpha$-th energy state which can be
extracted by using the eigen functions, $X$, in Eq.~(\ref{eq001}),
\begin{eqnarray}
|\Omega_\alpha\rangle = O_\beta^\dagger(\vec{0},0)|0>X_{\beta \alpha}.
\end{eqnarray}
We find that the wave functions are stable under the increase of the lattice volume and show clear characteristics 
of bound state for both channels, even for 
2S and 2P excited states. 

\section{Conclusion}

In order to study the charmonium dissociation at $T>0$,
we extract ground and the first excited charmonium states in S- and
P-waves and study boundary condition dependence of their energy, as
well as the volume-dependence of their wave functions. 
From a simulation of quenched anisotropic lattice QCD, 
we found so far no sign of scattering states up to about $2.3T_c$.

\section*{Acknowledgments}

The simulations have been performed on supercomputers (NEC SX-5) at
the Research Center for Nuclear Physics (RCNP) at Osaka University.
This work is also supported by the Large Scale Simulation Program
No.07-18(FY2008) of High Energy Accelerator Research Organization
(KEK) and the Nos.17340066 and Nos.19549001 with Grants-in-Aid of the
Japanese MEXT.


\begin{thebibliography}{99}

\bibitem{Umeda:2000ym}
T.~Umeda, R.~Katayama, O.~Miyamura and H.~Matsufuru,
\emph{Int.\ J.\ Mod.\ Phys.\ A} {\bf 16} (2001) 2215
[{\tt hep-lat/0011085}].

\bibitem{Umeda:2002vr}
T.~Umeda, K.~Nomura and H.~Matsufuru,
\emph{Eur.\ Phys.\ J.\ C} {\bf 39S1} (2005) 9
[{\tt hep-lat/0211003}].

\bibitem{Datta:2003ww}
S.~Datta, F.~Karsch, P.~Petreczky and I.~Wetzorke,
\emph{Phys.\ Rev.\ D} {\bf 69} (2004) 094507 
[{\tt hep-lat/0312037}].

\bibitem{Asakawa:2003re}
M.~Asakawa and T.~Hatsuda,
\emph{Phys.\ Rev.\ Lett.}  {\bf 92} (2004) 012001 
[{\tt hep-lat/0308034}].

\bibitem{Aarts:2006nr}
G.~Aarts, C.~Allton, M.~B.~Oktay, M.~Peardon and J.~I.~Skullerud,
  Phys.\ Rev.\  D {\bf 76}, 094513 (2007)
  [arXiv:0705.2198 [hep-lat]].

\bibitem{Jakovac:2006sf}
A.~Jakovac, P.~Petreczky, K.~Petrov and A.~Velytsky,
{\tt hep-lat/0611017}.

\bibitem{Digal:2001bh}
S.~Digal, P.~Petreczky and H.~Satz,
arXiv:hep-ph/0110406.

\bibitem{Karsch:2005nk}
F.~Karsch, D.~Kharzeev and H.~Satz,
Phys.\ Lett.\ B {\bf 637}, 75 (2006)
[arXiv:hep-ph/0512239].

\bibitem{Umeda:2007hy}
T.~Umeda,
\emph{Phys.\ Rev.\  D} {\bf 75} (2007) 094502
[{\tt hep-lat/0701005}].

\bibitem{Luscher:1990ck}
M.~Luscher and U.~Wolff,
Nucl.\ Phys.\  B {\bf 339}, 222 (1990).

\bibitem{Iida:2006mv}
H.~Iida, T.~Doi, N.~Ishii, H.~Suganuma and K.~Tsumura,
Phys.\ Rev.\ D {\bf 74}, 074502 (2006)
[arXiv:hep-lat/0602008].

\bibitem{Matsufuru:2001cp}
H.~Matsufuru, T.~Onogi and T.~Umeda,
\emph{Phys.\ Rev.\ D} {\bf 64} (2001) 114503
[{\tt hep-lat/0107001}].

\end{thebibliography}
\end{document}